\documentclass[a4paper]{article}

\usepackage{INTERSPEECH2021}
\usepackage{xcolor}
\usepackage{IEEEtrantools}  %
\usepackage{subfigure}
\usepackage[hyphens]{url}

\newcommand{\mypara}[1]{\vspace{2pt}\noindent{\bf{#1}}}

\usepackage[breaklinks=true,colorlinks,bookmarks=false]{hyperref}

\newcommand{\Audiocaps}{\textsc{AudioCaps}}
\newcommand{\Clotho}{\textsc{Clotho}}
\newcommand{\QuerYD}{\textsc{QuerYD}}

\newcommand{\activityNetLong}{\textsc{ActivityNet-Captions}}
\newcommand{\activityNetShort}{\textsc{ActNetCaps}}

\usepackage{bm}

\title{Audio Retrieval with Natural Language Queries}
\name{Andreea-Maria Oncescu$^{1*}$\thanks{$^*$ Authors contributed equally.}\quad \quad A. Sophia Koepke$^{2*}$ \quad \quad Jo{\~a}o F. Henriques$^1$\\ Zeynep Akata$^2$\quad \quad Samuel Albanie$^1$}

\address{
  \vspace{-0.25cm}
  $^1$VGG, University of Oxford \quad \quad
  $^2$EML, University of T{\"u}bingen\\
  {\normalsize \url{https://www.robots.ox.ac.uk/~vgg/research/audio-retrieval/}}
}  
\email{oncescu@robots.ox.ac.uk,  a-sophia.koepke@uni-tuebingen.de}

\begin{document}
\bstctlcite{IEEEexample:BSTcontrol}  %

\maketitle

\begin{abstract}
We consider the task of retrieving audio using free-form natural language queries. To study this problem, which has received limited attention in the existing literature, we introduce challenging new benchmarks for text-based audio retrieval using text annotations sourced from the \Audiocaps{} and \Clotho{} datasets.
We then employ these benchmarks to establish baselines for cross-modal audio retrieval, where we demonstrate the benefits of pre-training on diverse audio tasks.
We hope that our benchmarks will inspire further research into cross-modal text-based audio retrieval with free-form text queries.
\end{abstract}
\section{Introduction}
The large amounts of data being generated in recent years have prompted a real need to search evergrowing databases.
A search query typically comprises natural language (text), which allows expressing virtually any concept.
Spanning multiple modalities, different retrieval strategies were developed for content as diverse as text (including webpages and books), images~\cite{dong2016word2visualvec}, and videos \cite{miech2018learning,mithun2018learning}.
Surprisingly, while search engines currently exist for these modalities (e.g. Google, Flickr and YouTube, respectively), unstructured audio is not accessible in the same way.
The aim of this paper is to address this gap.

It is important to distinguish content-based retrieval from that based on meta-data, such as the title of a video or song or an audio tag.
Meta-data retrieval is feasible for manually-curated databases such as song or movie catalogs. However, content-based retrieval is more important in user-generated data, which often has little structure or meta-data.
There are methods to search for audio which matches an audio query \cite{Manocha18,lallemand2012content}, but satisfying the requirement to input an example audio query can be difficult for a human (e.g. making convincing frog sounds is difficult).
We, on the other hand, propose a framework which enables the querying of a sound database using detailed free-form natural language descriptions of the desired sound (e.g. ``A man talking as music is playing followed by a frog croaking.'').
This enables the retrieval of audio data which matches the temporal sequence of events in the query instead of just a single class tag.
Furthermore, natural language queries are a familiar user interface widely used in current search engines. Therefore, our proposed audio retrieval with free-form text queries could be a first step towards more natural and flexible audio-only search.

Text-based audio retrieval could also be beneficial for video retrieval. The majority of recent works that address the text-based video retrieval task focuses heavily on the visual and text domains~\cite{dong2016word2visualvec,miech2018learning,Liu19a,gabeur2020multi}. Since audio and visual information inherently have natural semantic alignment for a significant portion of video data, text-based audio retrieval could also be used for querying video databases by only considering the audio stream of the video data.
This would allow for video retrieval in the audio domain at reduced computational cost for cases in which audio and visual information correspond, with applications for low-power IoT devices, such as microphones in natural habitats, of particular interest for conservation and biology.
Historical archives with extensive sound collections, such as the BBC Sound Effects Archive, would be easier to search, facilitating historical research and public access.
Furthermore, text-based retrieval could enable appealing creative applications, such as automatically finding sounds which correspond to input text. This could be especially useful given the growing popularity of audio podcasts and audiobooks which are often supplemented with (background) sounds that match their content.

We propose two new benchmarks for text-based audio retrieval, based on the \Audiocaps{}~\cite{kim2019audiocaps} and \Clotho~\cite{drossos2020clotho} datasets. The former consists of a subset of 10-second audio clips from the AudioSet dataset~\cite{gemmeke2017audio} with additional human-written audio captions, while the latter contains audio captions for sounds sourced from the Freesound platform~\cite{font2013freesound}, varying between 15 and 30 seconds in duration.
In contrast to sound event class labels, audio captions contain detailed information about the sounds. A user searching for a particular sound would usually describe the sound using text similar to an audio caption. \Audiocaps{}~\cite{kim2019audiocaps} and \Clotho~\cite{drossos2020clotho} allow to leverage the matching audio-caption pairs to train text-based audio retrieval frameworks. To establish baselines for this task, we adapt existing video retrieval frameworks for audio retrieval. We employ multiple pre-trained audio expert networks and show that using an ensemble of audio experts improves audio retrieval.

In summary, we make three contributions:
(1) we introduce two new benchmarks for free-form text-based audio retrieval---to the best of our knowledge, these represent the first public benchmarks for this task;
(2) we provide baseline performances with existing multi-modal video retrieval models that we adapt to text-based audio retrieval;
(3) we demonstrate the benefits of combining multiple pre-training datasets.

\section{Related Work} \label{sec:related}

Our work relates to several themes in the literature: %
\textit{sound event recognition}, \textit{audio captioning}, \textit{audio-based retrieval}, \textit{text-based video retrieval} and \textit{text-domain audio retrieval}.
We discuss each of these next.

\mypara{Sound event recognition.}
The task of sound event recognition consists of matching audio information with a text label.
Research in this area has been driven by challenges, such as DCASE~\cite{stowell2015detection,mesaros2017dcase}, and by the collection of sound event datasets.
These include TUT Acoustic scenes \cite{mesaros_annamaria_2017_400515}, CHIME-Home~\cite{foster2015chime}, ESC-50~\cite{piczak2015esc},
\mbox{FSD}Kaggle~\cite{fonseca2019audio}, and AudioSet~\cite{gemmeke2017audio}.
Of relevance to our approach, a number of prior works have employed deep learning for audio comprehension~\cite{Kong18,yu2018multi,kong2019weakly,ford2019deep,kong2020panns} on the AudioSet dataset.
However, instead of considering the task of audio recognition, we focus on the task of retrieval with natural language queries.

\mypara{Audio captioning.}
Audio captioning consists of generating a text description for a sound~\cite{drossos2017automated}.
This requires a more detailed understanding of the sound than simply mapping the sound to a set of labels (sound event recognition).
Recently, several audio captioning datasets have been introduced, such as \Clotho{}~\cite{drossos2020clotho} which was used in the DCASE automated audio captioning challenge 2020~\cite{Dcase20}, Audio Caption \cite{wu2019audio},
and \Audiocaps{}~\cite{kim2019audiocaps}.
Multiple works have addressed automatic audio captioning on the \Audiocaps{} dataset \cite{koizumi2020audio,xu2021investigating,eren2020audio}.
In this work, we use the \Audiocaps{} and \Clotho{} datasets for cross-modal retrieval.

\mypara{Audio-based retrieval.}
Multiple content-based audio retrieval frameworks, in particular query by example methods, leverage the similarity of sound features that represent different aspects of sounds (e.g. pitch, or loudness) \cite{foote1997content, Wold96, helen2007query, lallemand2012content}.
More recently, \cite{Manocha18} use a siamese neural network framework to learn to encode semantically similar audio close together in the embedding space.
\cite{jin2012event} address multimedia event detection using only audio data,
while \cite{avgoustinakis2020audio} tackle near-duplicate video retrieval by audio retrieval.
These are purely audio-based methods that are applied to video datasets, but without using visual information. 
\cite{hou2013audio} propose a two-step approach for video retrieval which uses audio (coarse) and visual (fine) information together.

\mypara{Text-based video retrieval.}
More closely related to our work,
a number of methods showed that embedding video and text jointly into a shared space (such that their similarity can be computed efficiently) is an effective approach~\cite{dong2016word2visualvec,mithun2018learning,miech2018learning,Liu19a,wray2019fine,gabeur2020multi}. One particular trend has been to combine cues from several ``experts''---pre-trained models that specialise in different tasks (such as object recognition, action classification etc.) to inform the joint embedding.
In this work we propose to adapt two such methods: the Mixture of Embedded Experts method of~\cite{miech2018learning} and the Collaborative Experts model of~\cite{Liu19a}, by repurposing them for the task of audio retrieval (described in more detail in Sec.~\ref{sec:method}).

\mypara{Text-domain audio retrieval.}
Methods that retrieve audio by matching associated text, such as meta-data or sound event labels, have the implicit assumption that the text is relevant \cite{elizalde2018nels}.
\cite{slaney2002semantic} is an early work which
links audio and text representations in hierarchical semantic and acoustic spaces. \cite{chechik2008large} propose a text-based sound retrieval framework which uses single-word audio tags as queries rather than caption-like natural language.
The creative approach of \cite{aytar2017see} learns to align visual, audio, and text representations to enable cross-modal retrieval.
Their framework is trained with captioned images and paired image-sound data (sourced from videos) and evaluated using the soundtrack of captioned videos.
More recently,
\cite{elizalde2019cross} use a siamese network to learn a shared latent text and sound space for cross-modal retrieval.
While they use class labels as text labels, we study unconstrained text descriptions as queries. 
\section{Methods, Datasets and Benchmarks} \label{sec:method}

In this section, we first formulate the problem of audio retrieval with natural language queries.  Next, we describe two cross-modal embedding methods that we adapt for the task of audio retrieval. Finally, we describe the four datasets used in our experimental study, and the two benchmarks that we propose to evaluate performance on the audio retrieval task.

\mypara{Problem formulation.} Given a natural language query (i.e. a written description of an audio event to be retrieved) and a pool of audio samples, the objective of text-to-audio (abbreviated to \texttt{t2a}) retrieval is to rank the audio samples according to their similarity to the query. We also consider the converse \texttt{a2t} task, viz. retrieving text with audio queries.

\mypara{Methods.}
To tackle the problem of audio retrieval, we propose to learn cross-modal embeddings. Specifically, given a collection of $N$ audio samples with corresponding text descriptions, $\{(a_i, t_i) : i \in \{1, \dots, N\}\}$ we aim to learn embedding functions, $\psi_a$ and $\psi_t$, that project each audio sample $a_i$ and text sample $t_i$ into a shared space, such that $\psi_a(a_i)$ and $\psi_t(t_i)$ are close when the text describes the audio, and far apart otherwise. Writing $s_{ij}$ for the cosine similarity of the audio embedding $\psi_a(a_i)$ and the text embedding $\psi_t(t_j)$, we learn the embedding functions by minimising a contrastive ranking loss~\cite{socher2014grounded}:
\begin{align}
    \mathcal{L} = \frac{1}{B} \sum_{i=1, j \neq i}^B [m + s_{ij} - s_{ii}]_+ + [m + s_{ji} - s_{ii}]_+
    \label{eqn:loss}
\end{align}
where $B$ denotes the batch size, $m$ the \textit{margin} (set as a hyperparameter), and $[\cdot]_+ = max(\cdot, 0)$ the hinge function.

We consider two frameworks for learning such embedding functions $\psi_a$ and $\psi_t$: Mixture-of-Embedded Experts (MoEE)~\cite{miech2018learning} and Collaborative-Experts (CE)~\cite{Liu19a}.  MoEE, originally designed for video-text retrieval,  constructs its video encoder from a collection of ``experts'' (features extracted from networks pre-trained for object recognition, action classification, sound classification, etc.) which are computed for each video, aggregated along their temporal dimension with NetVLAD~\cite{arandjelovic2016netvlad}, projected to a lower dimension via a self-gated linear map and then L2-normalised.  The text encoder first embeds each word token with word2vec~\cite{mikolov2013efficient} and aggregates the results with NetVLAD~\cite{arandjelovic2016netvlad}. The result is projected by a sequence of self-gated linear maps (one for each expert) into a shared embedding space with the outputs of the video encoder. Finally, a scalar-weighted sum of the embedded experts in each joint space is used to compute the overall cosine similarity between the video and text (see~\cite{miech2018learning} for more details). CE adopts the same text encoder as MoEE and similarly makes use of multiple video experts. However, rather than projecting them directly into independent spaces against the embedded text, CE first applies a ``collaborative gating'' mechanism, which filters each expert with an element-wise attention mask that is generated with a small MLP that ingests all pairwise combinations of experts (see~\cite{Liu19a} for further details).

To adapt these frameworks for audio retrieval, we use the same text encoder structures $\psi_t$, and build audio encoders $\psi_a$ by mimicking the structure of their video encoders, replacing the ``video experts'' with ``audio experts'' (described in Sec.~\ref{sec:experiments}).

\mypara{Datasets.}
As the primary focus of our work, we study two \textit{audio-centric datasets}---these are datasets which comprise audio streams (sometimes with accompanying visual streams) paired with natural language descriptions that focus explicitly on the content of the audio track. To explore differences between audio retrieval and video retrieval, we also consider two \textit{visual-centric datasets}, that comprise audio and video streams paired with natural language which focus primarily (though not always exclusively) on the content of the video stream. Details of the four datasets we employ are given next. 

\noindent 1. \Audiocaps{}~\cite{kim2019audiocaps} \textit{(audio-centric)} is a dataset of sounds with event descriptions that was introduced for the task of audio captioning, with sounds sourced from the AudioSet dataset~\cite{gemmeke2017audio}.
Annotators were provided the audio tracks together with category hints (and with additional video hints if needed).
We use a subset of the data, excluding a small number of samples for which either: (i) the YouTube-hosted source video is no longer available,  (ii) the source video overlaps with the training partition of the VGGSound dataset~\cite{Chen20}.
Filtering to exclude samples affected by either issue leads to a dataset with 49,291 training, 428 validation and 816 test samples.\footnote{We make the sample list publicly available at the project page~\cite{projectPage}.}

\noindent 2. \Clotho{}~\cite{drossos2020clotho} \textit{(audio-centric)} is a dataset of described sounds that was also introduced for the task of audio captioning, with sounds sourced from the Freesound platform~\cite{font2013freesound}.
During labelling, annotators only had access to the audio stream (i.e. no visual stream or meta tags) to avoid their reliance on contextual information for removing ambiguity that could not be resolved from the audio stream alone.
The descriptions are filtered to exclude transcribed speech.
The publicly available version of the dataset includes a \texttt{dev} set of 2893 audio samples and an \texttt{evaluation} set of 1045 audio samples.
Every audio sample is accompanied by 5 written descriptions.
We used a random split of the \texttt{dev} set into a training and validation set with 2,314 and 579 samples, respectively.

\noindent 3. \activityNetLong{}~\cite{krishna2017dense} (visual-centric) consists of videos sourced
from YouTube and annotated with dense event descriptions.
It allocates 10,009 videos for training and 4,917 videos for testing (we use the public \texttt{val\_1} split provided by~\cite{krishna2017dense}).
For this dataset, descriptions also tend to focus on the visual stream.

\begin{table}[t]
\centering
\footnotesize
\setlength{\tabcolsep}{6pt}
\resizebox{\columnwidth}{!}{%
\begin{tabular}{l@{\hskip 0.15cm}c@{\hskip 0.2cm}c | @{\hskip 0.1cm}c@{\hskip 0.1cm}c|@{\hskip 0.1cm}c @{\hskip 0.1cm}c}
\hline \hline
\multicolumn{3}{c}{} &
\multicolumn{2}{c}{\hspace{-0.2cm}Text $\implies$ Audio} & \multicolumn{2}{c}{\hspace{-0.2cm}Audio $\implies$ Text} \\
Dataset & Anno. Focus & Pool & R$@$1 $\uparrow$ & R$@$10 $\uparrow$  & R$@$1 $\uparrow$ & R$@$10 $\uparrow$\\
\hline
\Audiocaps~\cite{kim2019audiocaps} & audio & 816 & 
$18.0_{\pm0.2}$&$62.0_{\pm0.5}$&$ 21.0_{\pm0.8} $&$62.7_{\pm1.6}$ \\
\Clotho~\cite{drossos2020clotho} & audio & 1045 & $4.0_{\pm0.2}$&$25.4_{\pm0.5}$&$ 4.8_{\pm0.4} $&$25.8_{\pm1.7}$\\
\hline
\activityNetShort{}~\cite{krishna2017dense} & visual & 4917 &$1.5_{\pm0.1}$&$9.2_{\pm0.3}$&$ 1.4_{\pm0.1} $&$8.5_{\pm0.3}$ \\
\QuerYD~\cite{Oncescu2021} & visual & 1954 & $3.7_{\pm0.2}$&$17.3_{\pm0.6}$&$ 3.8_{\pm0.2} $&$16.8_{\pm0.2}$ \\

\hline \hline
\end{tabular}
}
\caption{\textbf{Audio retrieval on audio-centric and visual-centric datasets.} Performance is strongest on the audio-centric \Audiocaps{} dataset and weakest on the visual-centric \activityNetLong{} (\activityNetShort) dataset.
}
\label{table:audio-retrieval-datasets}
\vspace{-0.9cm}
\end{table}

\noindent 4. \QuerYD{}~\cite{Oncescu2021} (visual-centric) is a dataset of described videos sourced from YouTube and the YouDescribe~\cite{youdescribe} platform.
It is accompanied by \textit{audio descriptions} that are provided with the explicit aim of conveying the video content to visually impaired users. Therefore, the provided descriptions focus heavily on the visual modality.
We use the version of the dataset comprising trimmed videos with 9,114 training, 1,952 validation, and 1,954 test samples.

\mypara{Benchmark.}
To facilitate the study of the text-based audio retrieval task, we propose to re-purpose the two \textit{audio-centric} datasets described above, \Audiocaps{} and \Clotho, to provide benchmarks for text-based audio retrieval. This approach is inspired by precedents in the vision and language communities, where datasets, such as~\cite{xu2016msr}, that were originally introduced for the task of video captioning, have become popular benchmarks for text-based video retrieval~\cite{miech2018learning,wray2019fine,Liu19a,gabeur2020multi}.

\begin{table}%
\centering
\footnotesize
\setlength{\tabcolsep}{6pt}
\resizebox{\columnwidth}{!}{%
\begin{tabular}{l | @{\hskip 0.1cm}c@{\hskip 0.1cm}c|@{\hskip 0.1cm}c @{\hskip 0.1cm}c}
\hline \hline
\multicolumn{1}{c}{} &
\multicolumn{2}{c}{\hspace{-0.2cm}Text $\implies$ Audio/Video} & \multicolumn{2}{c}{\hspace{-0.2cm}Audio/Video $\implies$ Text} \\
Expert & R$@$1  $\uparrow$ & R$@$10 $\uparrow$  & R$@$1 $\uparrow$ & R$@$10 $\uparrow$\\
\hline
\textbf{Visual experts only} &&&&\\
Scene & $6.1_{\pm0.4}$&$35.8_{\pm0.6}$& $6.5_{\pm0.8}$&$31.3_{\pm1.6}$ \\
Inst & $7.7_{\pm0.2}$&$46.7_{\pm1.3}$&$9.8_{\pm0.9} $&$40.6_{\pm0.7}$\\
R2P1D & $8.2_{\pm0.5}$&$44.7_{\pm0.9}$&$ 10.3_{\pm0.4} $&$41.8_{\pm3.1}$\\
Scene + Inst & $8.7_{\pm0.5}$&$47.4_{\pm0.5}$& $10.6_{\pm0.6} $&$41.4_{\pm1.5}$ \\
Scene + R2P1D & $8.8_{\pm0.1}$&$46.8_{\pm0.1}$&$ 11.0_{\pm0.6} $&$45.1_{\pm1.7}$ \\
R2P1D + Inst (\textit{CE-Visual}) & $ \bm{10.1_{\pm0.2}}$&$\bm{49.6_{\pm1.1}}$&$ \bm{12.1_{\pm0.4}} $&$\bm{46.1_{\pm1.3}}$ \\
\hline
\textbf{Audio experts only} &&&&\\
VGGish & $18.0_{\pm0.2}$&$62.0_{\pm0.5}$&$ 21.0_{\pm0.8} $&$62.7_{\pm1.6}$ \\
VGGSound & $20.5_{\pm0.6}$&$67.0_{\pm1.0}$&$ 24.6_{\pm0.9} $&$70.4_{\pm0.4}$\\
VGGish + VGGSound (\textit{CE-Audio}) & $\bm{23.1_{\pm0.8}}$&$\bm{70.7_{\pm0.7}}$&$ \bm{25.1_{\pm0.9}} $&$\bm{73.2_{\pm1.6}}$ \\
\hline
\textbf{Audio and visual experts} &&&&\\
\textit{CE-Visual} + VGGish & $23.9_{\pm0.7}$&$74.4_{\pm0.2}$&$ 29.0_{\pm2.0} $&$77.2_{\pm1.9}$\\
\textit{CE-Visual} + VGGSound & $27.4_{\pm0.7}$&$78.2_{\pm0.3}$&$ 34.0_{\pm1.5} $&$82.5_{\pm1.2}$ \\
\textit{CE-Visual} + \textit{CE-Audio}  & $\bm{28.1_{\pm0.6}}$&$\bm{79.0_{\pm0.5}}$&$ \bm{33.7_{\pm1.6}} $&$\bm{83.7_{\pm0.4}}$\\
\hline \hline
\end{tabular}
}
\caption{\textbf{The influence of different experts on \Audiocaps{}}. A comparison of audio and visual experts (applied to the video from which the audio was sourced) using CE~\cite{Liu19a}. We observe that audio features are significantly more effective than visual features (which nevertheless provide some complementary signal as can be seen when jointly using audio and visual features).
\vspace{-1.2cm}
}
\label{table:full-ce-audiocaps}
\end{table}

\section{Experiments} \label{sec:experiments}
\begin{figure*}[!ht]
\begin{minipage}[b]{0.79\linewidth}
\centering
\raisebox{0cm}{\includegraphics[width=\textwidth,clip,trim={0cm 0cm 0cm 0cm}]{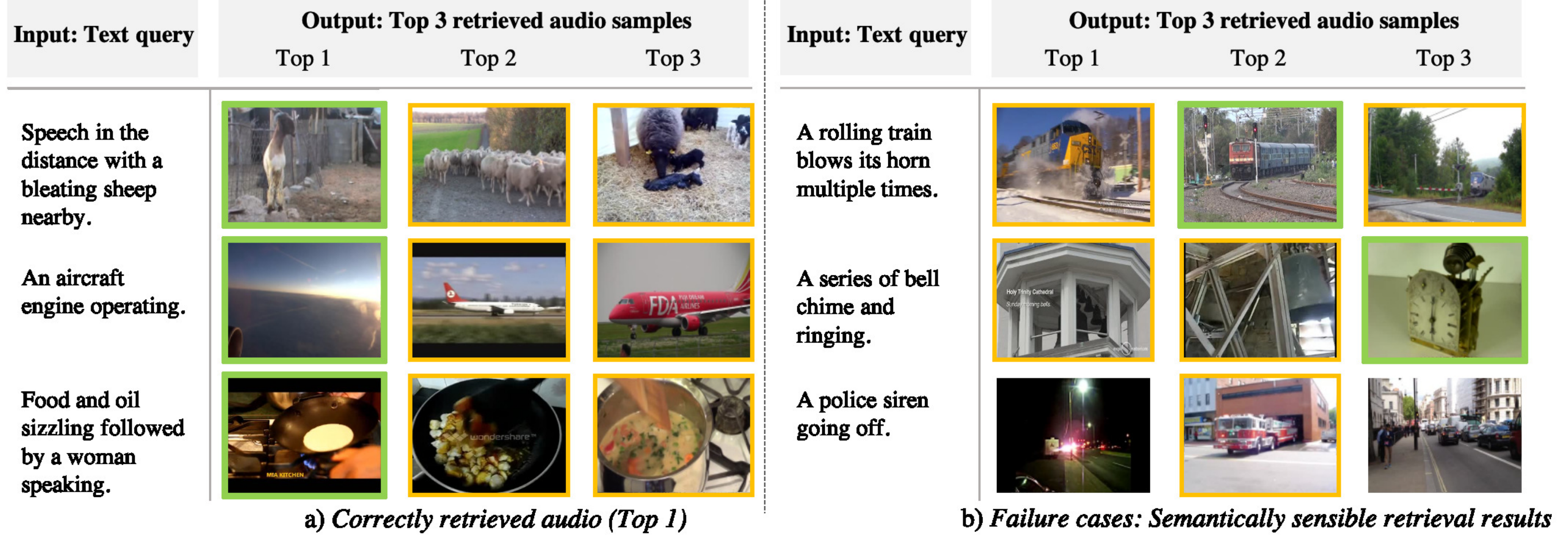}}
\caption{\textbf{Qualitative results.} Text-based audio retrieval results on \Audiocaps{} using CE with VGGish and VGGSound features. For an input text query, we visualise the top 3 retrieved audio samples using a video frame from the corresponding videos (the audio can be heard at the project webpage~\cite{projectPage}). We mark the audio samples which correspond to the query with green boxes. %
Successful retrievals are shown in a), failures in b). Note, in particular, the examples in b), where the model's top 1 retrieved audio is not the correct one, but the retrieved results nevertheless sound reasonable (visually convincing results are marked with yellow boxes).
}
\label{fig:qualitative}
\end{minipage}%
\hfill
\begin{minipage}[b]{0.195\linewidth}
\centering
  \raisebox{0cm}{\includegraphics[width=1\textwidth,clip,trim={0.2cm 0cm 0cm 0cm}]{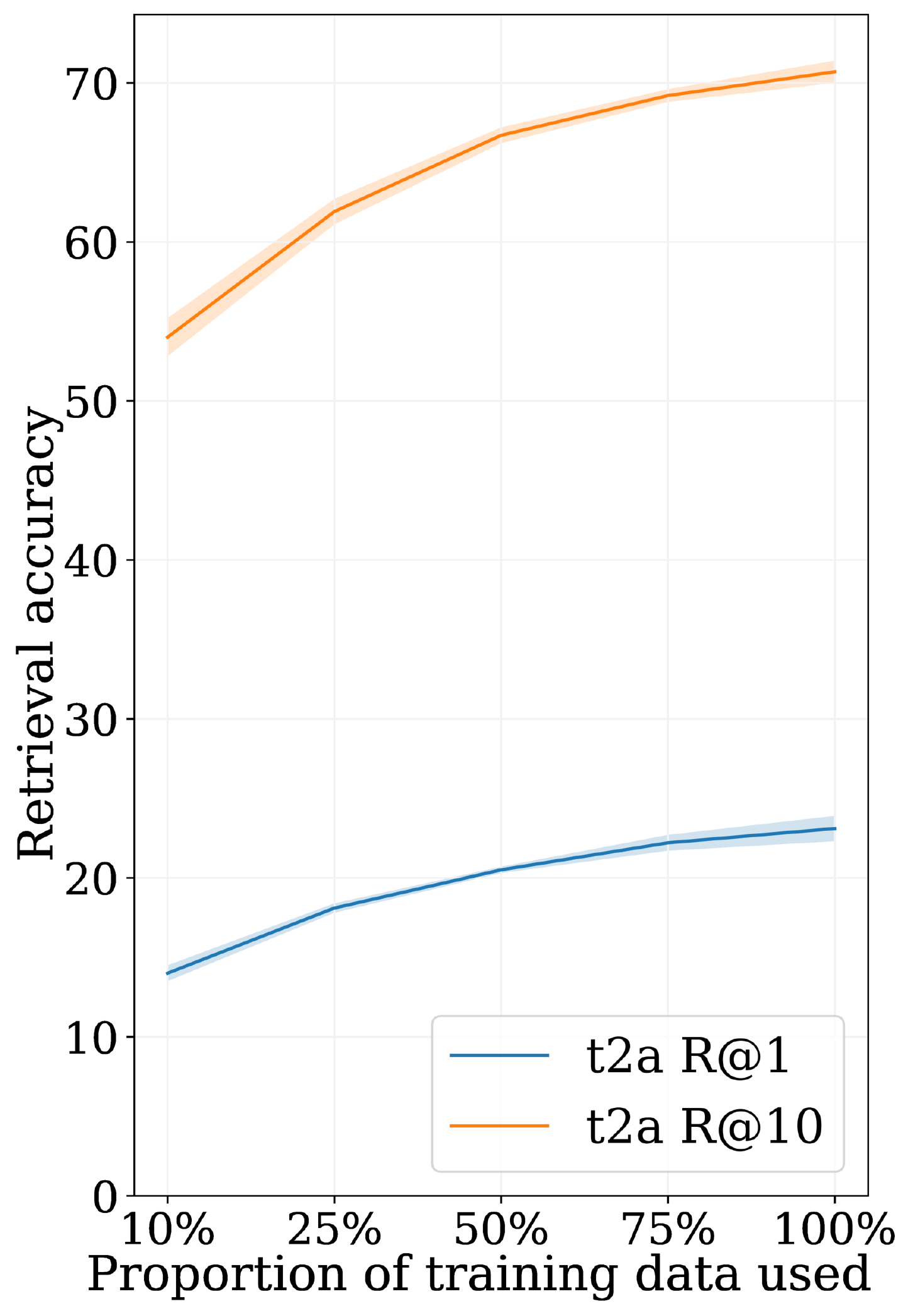}}
\caption{\textbf{The influence of training data scale.} CE retrieval performance on \Audiocaps{} with different proportions of the available \Audiocaps{} training data.}
\label{fig:trainset_ablation}
\end{minipage}
\end{figure*}

In this section, we first compare text-based audio retrieval and audio-based text retrieval performance on audio-centric and visual-centric datasets.  Next, we perform an ablation study on the contributions of different experts and present our baselines for the proposed \Audiocaps{} and \Clotho{} benchmarks. Finally we perform experiments to assess the influence of pre-training and training dataset size, and give qualitative examples of retrieval results.
Throughout the section, we use the standard retrieval metrics: recall at rank $k$ (R@$k$) which measures the percentage of targets retrieved within the top $k$ ranked results (higher is better) and report the mean and standard deviation of three different randomly seeded runs.\par

\mypara{Implementation details.} To encode the audio signal, we use two pre-trained audio feature extractors: \mbox{VGG}ish pre-trained for audio classification on YouTube8M~\cite{Hershey16}, and ResNet18 pre-trained for audio classification on the \mbox{VGG}Sound dataset~\cite{Chen20}.%
\footnote{Since the AudioCaps test set is a subset of the AudioSet training set (unbalanced), we do not use audio experts pre-trained on AudioSet.} All audio retrieval models were trained with a batch size of 128 using the contrastive ranking loss (Eqn.~\ref{eqn:loss}), with $m$ set to 0.2. The learning rate was set to 0.01 with a weight decay of 0.001.
Further  details  on  optimisation,  pre-trained experts (for audio and visual modalities) and hyperparameter choices are provided in the supplementary material.

\mypara{Audio-centric vs. visual-centric queries.} 
We first investigate audio retrieval with audio-centric and visual-centric queries. For this experiment, we use CE with a single expert (\mbox{VGG}ish audio features).
In Table~\ref{table:audio-retrieval-datasets}, we observe that performance is strongest on the audio-centric \Audiocaps{} dataset and weakest overall on the \activityNetLong{} dataset. We note that the \Clotho{} dataset is particularly challenging, with performance weaker (accounting for pool size) than the visual-centric \QuerYD{} dataset. We hypothesise that this is for two reasons: (1) the significantly smaller training set size of \Clotho, compared to all other datasets, (2) \Clotho{} was constructed by selecting audio samples that were as diverse as possible, making it a potentially more difficult benchmark.
We note, however, that the \QuerYD{} experiments suggest that computationally efficient video retrieval using only the audio stream can still be obtained, although at a lower accuracy.\par

\mypara{Ablation study.}
We next conduct an ablation study to investigate the effectiveness of different experts for audio retrieval on the \Audiocaps{} dataset.
We present the results in Tab.~\ref{table:full-ce-audiocaps}, where we observe that audio experts significantly outperform visual experts (pre-trained models for visual tasks like scene classification, which we compute from the video from which the audio was sourced). We note that the combination of audio and visual experts performs strongest overall, suggesting that the audio-centric queries contain information that is more accessible from the visual modality. The strongest audio-only retrieval is achieved by combining \mbox{VGG}ish and \mbox{VGG}Sound features---we therefore adopt this setting for the remaining experiments.

Additionally, we experimented with using speech features (word2vec~\cite{mikolov2013efficient} encodings of speech-to-text transcriptions~\cite{google-speech} of the audio stream). However, this did not improve the retrieval performance. Upon further investigation, we found that the audio captions and the spoken words in the \Audiocaps{} dataset do not have a significant overlap (corresponding to a METEOR~\cite{banerjee2005meteor} score of only 0.03).

\begin{table}
\vspace{-5pt}
\centering
\footnotesize
\setlength{\tabcolsep}{6pt}
\resizebox{\columnwidth}{!}{%
\begin{tabular}{l c |@{\hskip 0.2cm}c  @{\hskip 0.1cm}c | @{\hskip 0.1cm}c @{\hskip 0.1cm}c }
\hline \hline
\multicolumn{2}{c}{} &
\multicolumn{2}{c}{Text $\implies$ Audio} & \multicolumn{2}{c}{Audio $\implies$ Text} \\
Benchmark & Pre-training & R$@$1 $\uparrow$ & R$@$10 $\uparrow$  & R$@$1 $\uparrow$ & R$@$10 $\uparrow$\\
\hline
\textbf{\Audiocaps{}} &  &&&\\
MoEE & None & $22.5_{\pm0.3}$&$69.5_{\pm0.9}$&$ 25.1_{\pm0.8} $&$72.9_{\pm1.2}$ \\
CE & None & $23.1_{\pm0.8}$&$70.7_{\pm0.7}$&$ 25.1_{\pm0.9} $&$73.2_{\pm1.6}$ \\
\hline \hline
\textbf{\Clotho{}} &  &&&&\\
MoEE & None &$6.0_{\pm0.1}$&$32.3_{\pm0.3}$&$ 7.2_{\pm0.5} $&$33.2_{\pm1.1}$\\
CE & None &$6.7_{\pm0.4}$&$33.2_{\pm0.3}$&$ 7.1_{\pm0.3} $&$34.6_{\pm0.5}$\\
\hline
\textbf{\Clotho{}} &  &&&&\\
MoEE &\Audiocaps{}&$8.6_{\pm0.4}$&$39.3_{\pm0.7}$&$ 10.0_{\pm0.3} $&$40.1_{\pm1.3}$\\
CE &\Audiocaps{}& $9.6_{\pm0.3}$&$40.1_{\pm0.7}$&$ 10.7_{\pm0.6} $&$40.8_{\pm1.4}$\\
\hline \hline
\end{tabular}
}
\caption{\textbf{Audio retrieval benchmarks.} Text-audio and audio-text retrieval results for MoEE~\cite{miech2018learning} and CE~\cite{Liu19a} with VGGish and VGGSound features on the proposed \Audiocaps{} and \Clotho{} retrieval benchmarks. Pre-training on \Audiocaps{} improves the performance on \Clotho{}.
\vspace{-1cm}
}
\label{table:benchmarks}
\end{table}

\mypara{Benchmark results.} Incorporating the strongest combination of experts from the ablation study, we report our final baselines for text-audio and audio-text retrieval for two methods on the \Audiocaps{} and \Clotho{} datasets in Tab.~\ref{table:benchmarks}, where we observe that CE slightly outperforms MoEE. 
We also report the performance after pre-training models for retrieval on \Audiocaps{} and then fine-tuning on \Clotho{} (lower part of Tab.~\ref{table:benchmarks}). Here, we observe that pre-training on the audio-centric \Audiocaps{} brings a boost to both, CE and MoEE. 
Furthermore, we explored pre-training on \Clotho{} and fine-tuning on the \Audiocaps{} dataset, but found negligible change in performance (likely due to the fact that the \Audiocaps{} training set is significantly larger than that of \Clotho).

\mypara{Qualitative results.}  The qualitative results in Figure \ref{fig:qualitative} show examples in which the CE model with VGGish and VGGSound expert modalities (CE-Audio) is used to retrieve audio with natural language queries. The retrieved results mostly contain audio that is semantically similar to the input text queries. Observed failure cases arise from audio samples sounding very similar to one another despite being semantically distinct (e.g. the siren of a fire engine sounds very similar to a police siren).  

\mypara{Influence of scale.}
Finally, we present experiments using different proportions of the \Audiocaps{} dataset for training CE-Audio in Fig.~\ref{fig:trainset_ablation}. We observe that as more training data becomes available, retrieval performance increases monotonically. We also observe that there is still clear room for improvement in terms of retrieval results simply by collecting additional training data, motivating further dataset construction work to support future research on this important task.\par
\section{Conclusion} \label{sec:conclusion}
We introduced a new benchmark for natural language based audio retrieval, and provided baseline results by adapting strong multi-modal video retrieval methods.
Our results show that these methods are relatively well-suited for the audio retrieval task, however there is room for improvement, as expected for an under-explored problem.
We hope that our proposed benchmarks will facilitate the development of future audio search engines, and make this large fraction of the world's produced media available for public use.

\mypara{Acknowledgements.}
AMO was supported by an EPSRC DTA Studentship.
ASK and ZA were supported by the ERC (853489 - DEXIM) and by the DFG (2064/1 – Project number 390727645).
JFH is supported by the Royal Academy of Engineering (RF\textbackslash201819\textbackslash18\textbackslash163).
SA was supported by EPSRC EP/T028572/1 Visual AI.

\bibliographystyle{IEEEtran}
\bibliography{vgg_bibtex/shortstrings,vgg_bibtex/vgg_local,vgg_bibtex/vgg_other,mybib}

\begin{thebibliography}{10}
\providecommand{\url}[1]{#1}
\csname url@samestyle\endcsname
\providecommand{\newblock}{\relax}
\providecommand{\bibinfo}[2]{#2}
\providecommand{\BIBentrySTDinterwordspacing}{\spaceskip=0pt\relax}
\providecommand{\BIBentryALTinterwordstretchfactor}{4}
\providecommand{\BIBentryALTinterwordspacing}{\spaceskip=\fontdimen2\font plus
\BIBentryALTinterwordstretchfactor\fontdimen3\font minus
  \fontdimen4\font\relax}
\providecommand{\BIBforeignlanguage}[2]{{%
\expandafter\ifx\csname l@#1\endcsname\relax
\typeout{** WARNING: IEEEtran.bst: No hyphenation pattern has been}%
\typeout{** loaded for the language `#1'. Using the pattern for}%
\typeout{** the default language instead.}%
\else
\language=\csname l@#1\endcsname
\fi
#2}}
\providecommand{\BIBdecl}{\relax}
\BIBdecl

\bibitem{dong2016word2visualvec}
J.~Dong \emph{et~al.}, ``Word2visualvec: Image and video to sentence matching
  by visual feature prediction,'' \emph{arXiv:1604.06838}, 2016.

\bibitem{miech2018learning}
A.~Miech \emph{et~al.}, ``Learning a text-video embedding from incomplete and
  heterogeneous data,'' \emph{arXiv:1804.02516}, 2018.

\bibitem{mithun2018learning}
N.~C. Mithun \emph{et~al.}, ``Learning joint embedding with multimodal cues for
  cross-modal video-text retrieval,'' in \emph{Proc. ACM ICMR}, 2018.

\bibitem{Manocha18}
P.~Manocha \emph{et~al.}, ``Content-based representations of audio using
  siamese neural networks,'' in \emph{Proc. ICASSP}, 2018.

\bibitem{lallemand2012content}
I.~Lallemand \emph{et~al.}, ``Content-based retrieval of environmental sounds
  by multiresolution analysis,'' in \emph{Proc. SMC}, 2012.

\bibitem{Liu19a}
Y.~Liu \emph{et~al.}, ``Use what you have: Video retrieval using
  representations from collaborative experts,'' in \emph{Proc. BMVC}, 2019.

\bibitem{gabeur2020multi}
V.~Gabeur \emph{et~al.}, ``Multi-modal transformer for video retrieval,'' in
  \emph{Proc. ECCV}, 2020.

\bibitem{kim2019audiocaps}
C.~D. Kim \emph{et~al.}, ``Audiocaps: Generating captions for audios in the
  wild,'' in \emph{Proc. NACCL}, 2019.

\bibitem{drossos2020clotho}
K.~Drossos \emph{et~al.}, ``Clotho: An audio captioning dataset,'' in
  \emph{Proc. ICASSP}, 2020.

\bibitem{gemmeke2017audio}
J.~F. Gemmeke \emph{et~al.}, ``Audio set: An ontology and human-labeled dataset
  for audio events,'' in \emph{Proc. ICASSP}, 2017.

\bibitem{font2013freesound}
F.~Font \emph{et~al.}, ``Freesound technical demo,'' in \emph{Proc. ACM
  Multimedia}, 2013.

\bibitem{stowell2015detection}
D.~Stowell \emph{et~al.}, ``Detection and classification of acoustic scenes and
  events,'' \emph{IEEE Transactions on Multimedia}, 2015.

\bibitem{mesaros2017dcase}
A.~Mesaros \emph{et~al.}, ``{DCASE 2017 challenge setup: Tasks, datasets and
  baseline system},'' in \emph{DCASE 2017}, 2017.

\bibitem{mesaros_annamaria_2017_400515}
\BIBentryALTinterwordspacing
{A. Mesaros \etal}, ``{TUT Acoustic scenes 2017, Dev. dataset},'' Mar 2017.
  [Online]. Available: \url{https://doi.org/10.5281/zenodo.400515}
\BIBentrySTDinterwordspacing

\bibitem{foster2015chime}
P.~Foster \emph{et~al.}, ``Chime-home: A dataset for sound source recognition
  in a domestic environment,'' in \emph{IEEE WASPAA}, 2015.

\bibitem{piczak2015esc}
K.~J. Piczak, ``{ESC: Dataset for environmental sound classification},'' in
  \emph{Proc. ACM Multimedia}, 2015.

\bibitem{fonseca2019audio}
E.~Fonseca \emph{et~al.}, ``Audio tagging with noisy labels and minimal
  supervision,'' \emph{arXiv:1906.02975}, 2019.

\bibitem{Kong18}
Q.~Kong \emph{et~al.}, ``Audio set classification with attention model: A
  probabilistic perspective,'' in \emph{Proc. ICASSP}, 2018.

\bibitem{yu2018multi}
C.~Yu \emph{et~al.}, ``Multi-level attention model for weakly supervised audio
  classification,'' \emph{arXiv:1803.02353}, 2018.

\bibitem{kong2019weakly}
Q.~Kong \emph{et~al.}, ``Weakly labelled audioset tagging with attention neural
  networks,'' \emph{IEEE/ACM TASLP}, 2019.

\bibitem{ford2019deep}
L.~Ford \emph{et~al.}, ``A deep residual network for large-scale acoustic scene
  analysis.'' in \emph{INTERSPEECH}, 2019.

\bibitem{kong2020panns}
Q.~Kong \emph{et~al.}, ``Panns: Large-scale pretrained audio neural networks
  for audio pattern recognition,'' \emph{IEEE/ACM TASLP}, 2020.

\bibitem{drossos2017automated}
K.~Drossos \emph{et~al.}, ``Automated audio captioning with recurrent neural
  networks,'' in \emph{IEEE WASPAA}, 2017.

\bibitem{Dcase20}
\BIBentryALTinterwordspacing
``{DCASE2020} challenge task 6: Automated audio captioning,'' 2020. [Online].
  Available:
  \url{http://dcase.community/challenge2020/task-automatic-audio-captioning}
\BIBentrySTDinterwordspacing

\bibitem{wu2019audio}
M.~Wu \emph{et~al.}, ``Audio caption: Listen and tell,'' in \emph{Proc.
  ICASSP}, 2019.

\bibitem{koizumi2020audio}
Y.~Koizumi \emph{et~al.}, ``Audio captioning using pre-trained large-scale
  language model guided by audio-based similar caption retrieval,''
  \emph{arXiv:2012.07331}, 2020.

\bibitem{xu2021investigating}
X.~Xu \emph{et~al.}, ``Investigating local and global information for automated
  audio captioning with transfer learning,'' \emph{arXiv:2102.11457}, 2021.

\bibitem{eren2020audio}
A.~{\"O}. Eren and M.~Sert, ``Audio captioning based on combined audio and
  semantic embeddings,'' in \emph{IEEE ISM}, 2020.

\bibitem{foote1997content}
J.~T. Foote, ``Content-based retrieval of music and audio,'' in
  \emph{Multimedia Storage and Archiving Systems II}.\hskip 1em plus 0.5em
  minus 0.4em\relax International Society for Optics and Photonics, 1997.

\bibitem{Wold96}
E.~Wold \emph{et~al.}, ``Content-based classification, search, and retrieval of
  audio,'' \emph{IEEE Multimedia}, 1996.

\bibitem{helen2007query}
M.~Hel{\'e}n and T.~Virtanen, ``Query by example of audio signals using
  euclidean distance between gaussian mixture models,'' in \emph{Proc. ICASSP},
  2007.

\bibitem{jin2012event}
Q.~Jin \emph{et~al.}, ``Event-based video retrieval using audio,'' in
  \emph{Proc. ISCA}, 2012.

\bibitem{avgoustinakis2020audio}
P.~Avgoustinakis \emph{et~al.}, ``Audio-based near-duplicate video retrieval
  with audio similarity learning,'' \emph{arXiv:2010.08737}, 2020.

\bibitem{hou2013audio}
S.~Hou and S.~Zhou, ``Audio-visual-based query by example video retrieval,''
  \emph{Mathematical Problems in Engineering}, 2013.

\bibitem{wray2019fine}
M.~Wray \emph{et~al.}, ``Fine-grained action retrieval through multiple
  parts-of-speech embeddings,'' in \emph{Proc. ICCV}, 2019.

\bibitem{elizalde2018nels}
B.~Elizalde \emph{et~al.}, ``Nels-never-ending learner of sounds,''
  \emph{arXiv:1801.05544}, 2018.

\bibitem{slaney2002semantic}
M.~Slaney, ``Semantic-audio retrieval,'' in \emph{Proc. ICASSP}, 2002.

\bibitem{chechik2008large}
G.~Chechik \emph{et~al.}, ``Large-scale content-based audio retrieval from text
  queries,'' in \emph{Proc. ACM ICMIR}, 2008.

\bibitem{aytar2017see}
Y.~Aytar \emph{et~al.}, ``See, hear, and read: Deep aligned representations,''
  \emph{arXiv:1706.00932}, 2017.

\bibitem{elizalde2019cross}
B.~Elizalde \emph{et~al.}, ``Cross modal audio search and retrieval with joint
  embeddings based on text and audio,'' in \emph{Proc. ICASSP}, 2019.

\bibitem{socher2014grounded}
R.~Socher \emph{et~al.}, ``Grounded compositional semantics for finding and
  describing images with sentences,'' \emph{Transactions of the Association for
  Computational Linguistics}, 2014.

\bibitem{arandjelovic2016netvlad}
R.~Arandjelovic \emph{et~al.}, ``Netvlad: Cnn architecture for weakly
  supervised place recognition,'' in \emph{Proc. CVPR}, 2016.

\bibitem{mikolov2013efficient}
T.~Mikolov \emph{et~al.}, ``Efficient estimation of word representations in
  vector space,'' \emph{arXiv:1301.3781}, 2013.

\bibitem{Chen20}
H.~Chen \emph{et~al.}, ``Vggsound: A large-scale audio-visual dataset,'' in
  \emph{Proc. ICASSP}, 2020.

\bibitem{projectPage}
\BIBentryALTinterwordspacing
``Project page.'' [Online]. Available:
  \url{https://www.robots.ox.ac.uk/~vgg/research/audio-retrieval/}
\BIBentrySTDinterwordspacing

\bibitem{krishna2017dense}
R.~Krishna \emph{et~al.}, ``Dense-captioning events in videos,'' in \emph{Proc.
  ICCV}, 2017.

\bibitem{Oncescu2021}
A.-M. Oncescu \emph{et~al.}, ``{QuerYD: A video dataset with high-quality
  textual and audio narrations},'' in \emph{Proc. ICASSP}, 2021.

\bibitem{youdescribe}
\BIBentryALTinterwordspacing
{Video Description Research and Development Center}, ``Youdescribe,'' 2013.
  [Online]. Available: \url{http://youdescribe.ski.org}
\BIBentrySTDinterwordspacing

\bibitem{xu2016msr}
J.~Xu \emph{et~al.}, ``Msr-vtt: A large video description dataset for bridging
  video and language,'' in \emph{Proc. CVPR}, 2016.

\bibitem{Hershey16}
S.~Hershey \emph{et~al.}, ``{CNN} architectures for large-scale audio
  classification,'' in \emph{Proc. ICASSP}, 2016.

\bibitem{google-speech}
Google, ``{Speech-to-Text API},''
  \url{https://cloud.google.com/speech-to-text}.

\bibitem{banerjee2005meteor}
S.~Banerjee and A.~Lavie, ``{METEOR: An automatic metric for MT evaluation with
  improved correlation with human judgments},'' in \emph{Proc. ACL Workshop},
  2005.

\bibitem{abu2016youtube}
S.~Abu-El-Haija \emph{et~al.}, ``Youtube-8m: A large-scale video classification
  benchmark,'' \emph{arXiv preprint arXiv:1609.08675}, 2016.

\bibitem{He15}
K.~He \emph{et~al.}, ``Deep residual learning for image recognition,''
  \emph{arXiv preprint arXiv:1512.03385}, 2015.

\bibitem{xu2015jointly}
R.~Xu \emph{et~al.}, ``Jointly modeling deep video and compositional text to
  bridge vision and language in a unified framework,'' in \emph{AAAI}, 2015.

\bibitem{mahajan2018exploring}
D.~Mahajan \emph{et~al.}, ``Exploring the limits of weakly supervised
  pretraining,'' in \emph{Proc. ECCV}, 2018.

\bibitem{Deng09}
J.~Deng \emph{et~al.}, ``Imagenet: A large-scale hierarchical image database,''
  in \emph{Proc. CVPR}, 2009.

\bibitem{huang2017densely}
G.~Huang \emph{et~al.}, ``Densely connected convolutional networks,'' in
  \emph{Proc. CVPR}, 2017.

\bibitem{zhou2017places}
B.~Zhou \emph{et~al.}, ``Places: A 10 million image database for scene
  recognition,'' \emph{IEEE PAMI}, 2017.

\bibitem{tran2018closer}
D.~Tran \emph{et~al.}, ``A closer look at spatiotemporal convolutions for
  action recognition,'' in \emph{Proc. CVPR}, 2018.

\bibitem{ghadiyaram2019large}
D.~Ghadiyaram \emph{et~al.}, ``Large-scale weakly-supervised pre-training for
  video action recognition,'' in \emph{Proc. CVPR}, 2019.

\bibitem{zhang2019lookahead}
M.~R. Zhang \emph{et~al.}, ``Lookahead optimizer: k steps forward, 1 step
  back,'' in \emph{NeurIPS}, 2019.

\bibitem{liu2019variance}
L.~Liu \emph{et~al.}, ``On the variance of the adaptive learning rate and
  beyond,'' \emph{arXiv preprint arXiv:1908.03265}, 2019.

\bibitem{lesswright}
\BIBentryALTinterwordspacing
``Ranger optimiser.'' [Online]. Available:
  \url{https://github.com/lessw2020/Ranger-Deep-Learning-Optimizer}
\BIBentrySTDinterwordspacing

\end{thebibliography}
\clearpage
\appendix
\begin{section}{Supplementary Material}
We give details on the pre-trained experts used, and on optimisation and hyperparameter choices below.

\begin{subsection}{Pre-trained experts}
This section contains information about the audio and visual expert features that were obtained from pre-trained networks.

\begin{subsubsection}{Audio experts}
We use two different audio experts, which we abbreviate as \mbox{VGG}ish and \mbox{VGG}Sound.

\mypara{\mbox{VGG}ish.}
Theses audio features are obtained with a \mbox{VGG}ish model \cite{Hershey16}, trained for audio classification on the YouTube-8M dataset \cite{abu2016youtube}. To produce the input for this model, the audio stream of each video is re-sampled to a 16kHz mono signal, converted to an STFT with a window size of 25ms and a hop size of 10ms with a Hann window, then mapped to a 64 bin log mel spectrogram. Finally, the features are parsed into non-overlapping 0.96s collections of frames (each collection comprises 96 frames, each of 10ms duration), which is mapped to a 128-dimensional feature vector.

\mypara{VGGSound.}
These features are extracted using a ResNet-18 model \cite{He15} that has been pre-trained on the \mbox{VGG}Sound dataset (model H) \cite{Chen20}. We modify the last average pooling layer to aggregate along the frequency dimension, but keep the full temporal dimension. This results in features of dimension t x 512, where t denotes the number of time steps.
\end{subsubsection}

\begin{subsubsection}{Visual experts}
We use three different visual experts, abbreviated as Inst, Scene, and R2P1D.

\mypara{Inst.}
These features are extracted using a ResNeXt-101 model \cite{xu2015jointly} that has been pre-trained on Instagram hashtags \cite{mahajan2018exploring} and finetuned on ImageNet \cite{Deng09} for the task of image classification. Features are extracted from frames extracted at 25 fps, where each frame is resized to 224 × 224 pixels. Embeddings are 2048-dimensional.

\mypara{Scene.}
Scene features are extracted from 224×224 pixel centre crops with a DenseNet-161 model \cite{huang2017densely} pre-trained on Places365 \cite{zhou2017places}. Embeddings are 2208-dimensional.

\mypara{R2P1D.}
Features are extracted with a 34-layer R(2+1)D model \cite{tran2018closer} trained on IG-65M \cite{ghadiyaram2019large} which processes clips of 8 consecutive 112 × 112 pixel frames, extracted at 30 fps. Embeddings are 512-dimensional.
\end{subsubsection}

\end{subsection}

\begin{subsection}{Optimisation details and hyperparameter choices}

\begin{subsubsection}{Optimiser}
We used the Lookahead solver \cite{zhang2019lookahead} in combination with RAdam \cite{liu2019variance} (implementation by \cite{lesswright}) with an intial learning rate of $0.01$ and weight decay of $0.001$. We use a learning rate decay for each parameter group with a factor of $0.95$ every epoch. Most models were trained for a maximum of 20 epochs whilst the Activity-Net model was trained for 40 epochs. We chose the models that gave the best performance on the geometric mean of R@1, R@5, and R@10.
\end{subsubsection}

\begin{subsubsection}{NetVLAD hyperparameters}
For text, we used 20 VLAD clusters and one ghost cluster. For the VGGish and VGGSound features we used 16 VLAD clusters.
\end{subsubsection}

\begin{subsubsection}{Zero-padding hyperparameters}
In the main paper, we used zero-padding of variable length experts to a common shape $z = (z_t,z_a,z_v)$. For \Audiocaps{} we have used $z_t=20$ for text tokens, $z_a=29$ for VGGish feature tokens, and $z_v=29$ for the VGGSound features. For \Clotho{} we have used zero-padding with $z_t=21$ for text tokens, $z_a=31$ for VGGish feature tokens, and $z_v=95$ for the VGGSound features.\par
For QuerYD we have used $z_t=70$ for text tokens and $z_a=500$ for VGGish feature tokens. Lastly, for Activity-Net we used $z_t=20$ for text tokens and $z_a=29$ for VGGish feature tokens. \par

\end{subsubsection}

\end{subsection}
\end{section}
\end{document}